\documentclass[aps,twocolumn,prd,superscriptaddress,preprintnumbers,showpacs]{revtex4}
\usepackage{graphicx}

\newcommand{\be}{\begin{equation}}
\newcommand{\ee}{\end{equation}}
\newcommand{\bea}{\begin{eqnarray}}
\newcommand{\eea}{\end{eqnarray}}

\newcommand{\lpl}{\ell_{\rm pl}}

\begin{document}

\preprint{FTPI-MINN-06/24}
\preprint{UMN-TH-2510/06}

\title{Constraints on a scale invariant power spectrum from superinflation in LQC}

\author{David J. Mulryne}
\email[Electronic address: ]{d.j.mulryne@qmul.ac.uk}
\affiliation{School of Mathematical Sciences, Queen Mary, University of London, Mile End Road, London E1 4NS, UK}
\author{Nelson J. Nunes}
\email[Electronic address: ]{nunes@physics.umn.edu}
\affiliation{School of Physics and Astronomy, University of Minnesota, 116 Church Street S.E., Minneapolis, Minnesota 55455, USA}

\date{\today}

\begin{abstract}
The computation of the spectrum of primordial perturbations, generated by a scalar field during the superinflationary phase of Loop Quantum Cosmology, is revisited. The calculation is performed for two different cases. The first considers the dynamics of a massless field and it is found that scale invariance can only be achieved under a severe fine tuning. The second assumes that the field evolves with a constant ratio between kinetic and potential energy, i.e. in a scaling solution. In this case, near scale invariance is a generic feature of the theory if the field rolls in a steep self interaction potential. 
\end{abstract}

\maketitle


\section{Introduction}

The inflationary scenario is currently the favored model
for the evolution of the very early universe \cite{Starobinsky:1980te,Guth:1980zm,Albrecht:1982wi,Hawking:1981fz,Linde:1981mu,Linde:1983gd}.
Inflation arises whenever the universe undergoes a phase of accelerated
expansion and was originally introduced to
solve a number
of perceived problems with the hot big bang model of the universe, including
the flatness, horizon and monopole problems \cite{Linde:1981mu}.
More importantly, however,
inflation is currently the favored model for large-scale structure
formation since it can create a scale invariant spectrum
of primordial density fluctuations, which provide the seeds of cosmic
structure \cite{Liddle:2000cg}. In the simplest versions of the scenario, 
inflation is realized by a scalar field, the inflaton, whose kinetic 
energy is negligible when compared to its potential energy such 
that $\dot{\phi}^2 < V$ \cite{Linde:1983gd}. This is typically called, 
slow-roll inflation.

Given the importance of inflation and that it occurred in the
early stages of the universe's evolution, in potentially high curvature
and density regimes, it is natural to investigate connections between
inflation and quantum cosmology.  This has recently been done in
the context of Loop Quantum Cosmology (LQC), which is the application of
Loop Quantum Gravity (LQG) to symmetric states (for reviews, see
Refs.~\cite{Rovelli:1997yv,Thiemann:2002nj,Bojowald:2005bm}).  
In particular LQC gives rise to a
``semiclassical'' regime in which the standard equations become
modified by non-perturbative quantum geometrical effects.
These semiclassical equations 
have been employed to study potential connections between
LQC and inflation, and a number of important
results have been obtained.
For example in the context of a universe
sourced by a minimally coupled scalar field, the
semiclassical modifications
cause an anti-frictional effect which accelerates the field
along its self-interaction
potential. In principle this effect can push the field up its potential and
set the initial conditions for subsequent slow-roll inflation
\cite{Bojowald:2004xq, Tsujikawa:2003vr,Lidsey:2004ef,
Mulryne:2004va,Mulryne:2005ef,Nunes:2005ra}.

The most striking feature of LQC, however, is
that the anti-frictional effect also causes the
universe to undergo an inflationary period \cite{Bojowald:2002nz}.
In contrast to standard slow-roll inflation, where inflation
is driven by the self-interaction potential of the scalar field,
in LQC the inflationary phase is now driven by quantum geometrical
effects.
Moreover, it is possible to
show that this period of inflation will occur independently of
any particular form of the potential \cite{hossain2}. LQC
therefore naturally predicts that the universe 
must evolve through an inflationary era, irrespective of whether
this era is followed by a phase of standard
slow-roll inflation or not.

It is natural to ask therefore, whether or not
this period of LQC inflation is able either to replace or to
supplement standard inflation, and what its observational
signatures would be.
In order to answer this question
one must consider both the number of e-folds of inflation
which the LQC phase can give rise to, and the spectrum of
perturbations which this phase will produce.

 The first of these issues has been addressed previously
\cite{Bojowald:2003mc}
and the conclusion was found to depend both on initial conditions
and the value of a particular quantization ambiguity parameter
labeled $j$ \cite{Bojowald:2002ny}. In order to solve the problems of the hot
big bang model, the required value of the parameter $j$ is very large.
That LQC inflation can replace standard inflation therefore seems
disfavored, given that smaller values of the parameter $j$ can
be argued to be more natural than larger ones \cite{Bojowald:2004xq}.

The issue concerning the spectrum of perturbations 
produced by the LQC inflationary phase, which might leave a 
signature of this phase, is a more subtle one.
An important
point is that during the LQC inflationary phase, not only does
the growth of the scale factor
accelerate, but the Hubble parameter also grows, $\dot{H}>0$.
Hence this phase is actually a superinflationary one, and experience from
standard inflation suggests that we should expect the spectrum of
perturbations to be strongly blue tilted ($n_s > 1$, where $n_s$ is the
spectral index).
A recent study finds, however, that the LQC inflationary
scenario can produce a nearly scale invariant spectrum of perturbations
\cite{Hossain:2004wm}. The study also finds that a generic prediction is that
the spectrum will be slightly blue tilted, in contrast to
most slow-roll models which have a small red tilt, and that this
result is robust, being independent of ambiguities in the quantization
scheme.
This might lead one to believe that near scale invariance with a
small blue tilt is
a generic and observationally falsifiable result of LQC,
in contrast with standard inflation
where there is a large amount of freedom in the value of the spectral
index associated with the form of the potential.
The calculation of the power spectrum in Ref.~\cite{Hossain:2004wm},
however, uses the so called
direct method \cite{Padmanabhan:1988se}.
This method is not the standard one which is normally invoked for
calculating the power spectrum of slow-roll inflation, but it is argued in Ref.~\cite{Hossain:2004wm} that the use of the direct method is
more natural within LQC because of the minimum natural length scale
introduced by LQC, and the lack of a general expression for the stress
energy tensor in LQC.

Two further important
aspects of the calculation in Ref.~\cite{Hossain:2004wm} are important to note.
First, it assumes that the effective equation of state,
$w= p/\rho$, for the universe as a whole is
given by $w \approx -1$.  This can only be true either at the
end of the superinflationary phase, or 
under severe fine tuning of the model's parameters.
Secondly, it assumes that
the background spacetime in which the scalar field lives is
unperturbed, and considers only perturbations in the
scalar field.  At the present time this is a necessary assumption, 
as the modified semiclassical equations of LQC
are only known for an unperturbed background.  
This assumption is technically invalid as
it clearly violates Einstein's field equations, however, whether
it proves to be a useful approximation remains to be seen.  
Experience from other applications
of perturbation theory in the early universe suggest that 
in some cases this approximation is very useful.
For example in standard slow-roll inflation a calculation 
of the spectrum of scalar field perturbations, using this approximation, 
can be applied to produce an accurate estimate for the 
spectrum of the comoving curvature perturbations, which is ultimately the 
important quantity for observations \cite{Liddle:2000cg,Dodelson:2003ft}.  
In other situations it is less 
useful, for example in the ekpyrotic scenario the 
application of a similar procedure to that used for slow-roll inflation 
yields erroneous predictions (see \cite{Khoury:2001wf,Lyth:2001pf}). 

In this short note, we readdress the question of whether a
scale invariant spectrum of primordial perturbations can be produced
during the LQC superinflationary era using methods
commonly employed in inflationary cosmology. Moreover, we do not make 
any assumptions about
the universe's expansion
rate, rather allowing it to be determined by the LQC dynamics.
We continue to use the approximation in which the background
space-time is unperturbed and therefore focus on the spectrum 
of scalar field perturbations produced in this approximation. 

We proceed as follows.  In section II we introduce the semiclassical regime, 
we summarize previous results which we require in section III, and our new results 
are presented in sections IV and V. Finally we conclude in section VI with a 
summary of what we have learned, and a discussion of how the results may be 
useful in the future. 

\section{Semi-classical Dynamics}
LQC is based on a Hamiltonian formulation of General Relativity.  
The dynamics is therefore governed by a Hamiltonian constraint equation 
which we can present schematically as 
${\cal H}_{\rm gravity}+ {\cal H}_{\rm matter}=0 \,,$
where we have indicated that the constraint consists 
of a gravitational and a matter part.  

The effective or semiclassical equations of LQC 
arise by incorporating into the classical Hamiltonian 
non-perturbative quantum effects from the underling LQG 
quantization procedure. A number of approaches have 
been taken to derive and verify 
the resulting effective Hamiltonian \cite{Date:2004zd, Bojowald:2004zf, Singh:2005xg, Vandersloot:2005kh}. A robust result of these approaches is the
introduction of modifications which come from the 
quantization of inverse metrical quantities.
In general 
such corrections occur both in the matter and the gravitational 
parts of the constraint, but most attention (with the exception of \cite{Vandersloot:2005kh}) 
has so far been focused 
on corrections to the matter Hamiltonian which give rise to the 
superinflationary effect.
Let us now consider this Hamiltonian in more detail when the 
matter source is a scalar field.

A crucial first step in formulating LQG is to rewrite 
canonical gravity in terms of Ashtekar variables, which are the 
densitised triad $E^a_i$ and the Ashtekar connection $A^i_a$ 
where $E^a_i = e^a_i/|\det e^b_i|$, with 
$e^a_i e^b_i = q^{ab}$ with 
$q_{ab}$ the spatial metric, 
and $A^i_a = \Gamma^i_a + K^i_a$ 
with 
$\Gamma$ the spin connection and $K$ the extrinsic curvature. The indices run from one to three.
When written in Ashtekar variables the matter Hamiltonian for a 
general spacetime becomes
\be
\label{HamPhi}
{\cal H}_\phi = \frac{p_\phi^2 }{2\sqrt{|\det E^c_j|}} +\frac{E^a_iE^b_i\partial_a\phi \partial_b\phi}{2\sqrt{|\det E^c_j|}} + \sqrt{|\det E^c_j|} \, V(\phi)\,~.
\ee

The terms in this Hamiltonian which involve inverse expressions 
cannot be quantized in a 
straight forward manner and must first be regularized by a 
procedure introduced by Thiemann \cite{Thiemann:1996aw,Thiemann:1997rt}.  
The expressions which 
result from this procedure are rather complicated, and in particular 
are subject to an number of quantization ambiguity parameters.

Here we are interested in isotopic LQC and in this case 
the matter Hamiltonian reduces to 
\be
\label{HamIso}
{\cal H}_\phi = \frac{1}{2}\frac{p_\phi^2}{\sqrt{|\det E^c_j|}} +\sqrt{|\det E^c_j|}  \, V(\phi)\,,
\ee
where we have set the gradient terms, which would violate isotropy, to zero. 
We are also assuming, for
simplicity, that $E$ represents the isotopic triad.  
In this setting the only inverse term is the inverse volume $(|\det E^c_j|^{-1/2}$), 
which classically is simply $a^{-3}$ in terms of metric variables.  Quantum mechanically 
however, this term 
must be quantized following Thiemann's prescription (for details see \cite{Bojowald:2001vw}).

The spectrum for the inverse volume can be calculated 
exactly in isotropic LQC, but because of the regularization   
the answer depends on a number of ambiguity parameters \cite{Bojowald:2002ny}. 
Above the scale of discreteness set by $a_i= \sqrt{\gamma} \lpl$, where $\gamma=0.27$ 
is the Barbero Immizi parameter, spacetime can be considered to be continuous, 
and the inverse spectrum can be approximated by a 
continuous function. There is a second scale of importance, however, which is 
set by a second quantity, $a_* = a_i \sqrt{j/3}$, 
where $j$, which takes half integer values, 
is one of the quantization ambiguities.  
Above this scale the eigenvalues of the inverse operator follow the 
classical values, while below it they are radically different.
In fact, where the classical inverse volume is infinite, i.e. at the classical 
singularity, in LQC the inverse volume is zero.  
We find therefore that 
if $a_*>a_i$, which implies $j>3$,  then there is an overlap 
between the regions in which the inverse volume can be approximated by a 
continuous function, and where the inverse volume deviates significantly 
from the classical expression.  This is the semiclassical regime, and 
the semiclassical matter Hamiltonian is  
simply arrived at by replacing the inverse volume 
term in Eq.~(\ref{HamIso}) by the continuous approximation function.

The function which approximates the spectrum of the 
inverse volume is given by $a^{-3} D(q)$, 
where $q \equiv (a/a_*)^2$ and
\begin{eqnarray} 
\label{defD}
D(q) &=&
\left\{\frac{3}{2l}q^{1-l}\left[(l+2)^{-1}
\left((q+1)^{l+2}-|q-1|^{l+2}\right)\right.\right. \nonumber\\
 &~& -\frac{1}{1 + l}q
\left((q+1)^{l+1}\right. \nonumber \\
 &~& \left.\left.\left. - {\rm sgn}(q-1) 
|q-1|^{l+1}\right)\right]\right\}^{3/(2-2l)} \,, 
\end{eqnarray}
where $l$ is another quantization ambiguity. From considerations of 
the regularization procedure within LQC, $l$ 
is constrained to the range $0<l<1$, 
while from considerations of the procedure within the full theory 
of LQG it must take a discrete set of values given by 
$l_k=1-(2k)^{-1}\geq1/2$, where $k$ is an integer \cite{Bojowald:2002ny}.
The expression for $D(q)$ for $a \ll a_*$ can be approximated by
$D(q) \approx
(3/(1+l))^{3/(2-2l)}q^{3(2-l)/2(1-l)} $.  It has a global
maximum at $a \approx a_*$ and falls monotonically to $D = 1$
for $a > a_*$.  Hence the classical inverse volume, $a^{-3}$
is recovered for $a \gg a_*$.

Replacing the inverse volume with this function one arrives at 
the semiclassical Hamiltonian, which we give here in terms of the 
scale factor including both the matter and gravitational parts: 
\be
\label{HamSC}
{\cal H}=-\frac{3}{8\pi \lpl^2} \, \dot{a}^2a + 
\frac{1}{2} Da^{-3} p_{\phi}^2 + a^3 V =0\,~.
\ee
We can now derive the semiclassical equations of motion.

Using this Hamiltonian density and considering  
$\dot{\phi}= \partial {\cal H} / \partial p_\phi $ 
we find that $p_{\phi}= a^3\dot{\phi}/D$, and hence
${\cal H}_{\phi} + a^3 \dot{\phi}^2/2D + a^3V$.  Then on
dividing the Hamiltonian constraint by $a^3$,
we arrive at the modified Friedmann equation
\be \label{friedeq}
H^2 \equiv \left(\frac{\dot{a}}{a}\right)^2 =
\frac{8\pi\lpl^2}{3} \, \left[ \frac{1}{2} \, D^{-1}\, \dot{\phi}^2 +
V(\phi)\right]\, ~.
\ee
The other dynamical equations can also be found using, 
$\dot{p}_\phi =-\partial {\cal H}/\partial \phi$,
 \be
\ddot{\phi} +3H\left(1 - \frac{1}{3} \,\frac{d\ln D}{d\ln a}\right) \dot{\phi}+ D \, \frac{d V(\phi)}{d\phi}
= 0\, ~,
\label{fieldeq}
 \ee
and combining Eq.~(\ref{fieldeq}) with (\ref{friedeq}) we also find
\be
\dot{H} =  -4\pi \lpl^2 \frac{\dot{\phi}^2}{D}
\left( 1-\frac{1}{6} \frac{d \ln D}{d \ln a}  \right) \,~.
\label{raycheq}
\ee

\section{Background evolution}
Before moving on to considering perturbations of the 
scalar field we require an understanding of the background dynamics.  
In particular we confine ourselves to the regime $a \ll a_*$, where 
analytic progress can be made. 

In this regime, we write the correction function as $D = D_* a^n$ with $n={3(2-l)/(1-l)}$ (hence $6<n<\infty$) and $D_*=(3/(1+l))^{3/(2-2l)}a_*^{3(l-2)/(1-l)}$. 
From Eq.~(\ref{raycheq}) it can be seen that the universe undergoes 
superinflationary expansion, $\dot{H}>0$, for $n > 6$, 
independently of the form of the self-interaction potential.
We will be interested in the cases where the ratio $\sqrt{2 D} \, H/\dot{\phi} = $ 
constant. 
This comprises the cases of a massless scalar field and of a scaling solution 
(when the ratio of kinetic to potential energy is a constant).
In both these cases, the
evolution can be solved exactly, and the scale factor undergoes 
power law growth.
When we come to deal with the perturbed equations, we 
will find it more convenient to
work with conformal time $dt=a \, d\tau$, and so we give the 
background evolution using this time variable.

\subsection{Massless scalar field}

Considering a massless scalar field, Eqs.~(\ref{friedeq}),
(\ref{fieldeq}) and (\ref{raycheq})
can be solved to yield
$a^{2-n/2} = a_{\rm init}^{2-n/2} + A^{2-n/2}(\tau_{\rm init}-\tau)$ where
\be
\label{eqA}
A = \left[ \left(\frac{n}{2}-2\right) \left(\frac{4\pi\lpl^2}{3} D_* \phi_{\rm init}^{'2} \frac{a_{\rm init}^4}{D_{\rm init}^2} \right)^{1/2} \right]^{2/(4-n)} \,,
\ee
which is positive-definite for $n > 4$.
It is convenient to rescale time to absorb the constant term into our definition of conformal time such that
\be
\label{masslesseq}
a = A \,(-\tau)^p \,,
\ee
where $p=2/(4-n) < 0$. It is worth pointing out that $\tau$ is negative and increasing for an expanding universe and decreasing for a contracting universe. 

\subsection{Scaling Solution}
A second way to achieve power law growth in the regime $a \ll a_*$ 
is for the field to roll in a self interactive potential 
of the form \cite{Lidsey:2004uz}
\begin{equation}
\label{potential}
V(\phi) = V_0 \, |\phi|^{\beta} \,,
\end{equation}
with $\beta > 0$.

The analogous scaling solution in the classical regime,
which is exponential in form, is very important in understanding 
the production and evolution of perturbations in the standard single field inflationary scenario, and we expect that
this scaling solution has the same degree of importance in the LQC scenario.

Using a rescaled conformal time, the growth of the scale factor and the
field are then determined by the expressions
\begin{eqnarray}
 a = A \, (-\tau)^p \,, \hspace{1cm} \phi = F \, (-\tau)^v \,, 
\end{eqnarray}
where $v = np/2$, $p = -4/(n\beta+4) <0$. $V_0$ is related to the 
constants $A$ and $F$ and the powers $v$, $n$ and $p$, however, 
this relation is not important in what follows. Obviously, the constant $A$, does not need to take the same value as in Eq.~(\ref{eqA}).

\section{Perturbation Theory}
If we are to fully understand the evolution of 
cosmological perturbations in LQC, we
must perturb both the gravitational and the matter 
sectors of the theory, about the homogeneous background.
So far, however, the quantization procedure in LQC has only 
been performed for homogeneous spacetimes, and not for 
the perturbed cases.  This means that
the full perturbed semiclassical equations have so far 
not been derived.  
In their absence, we may adopt a more modest
approach.  
This is, to assume that the background spacetime is
unperturbed, but to allow perturbations in the 
scalar field. The assumption is valid for cases in which perturbations of 
spacetime are much smaller than those of the matter source, or equivalently 
where the matter perturbations have a negligible effect on 
the background spacetime.
In this case we can calculate the power spectrum of the
scalar field's perturbations on super-horizon scales produced
from quantum mechanical fluctuations.  

Ultimately the quantity which is relevant to observations 
after the inflationary era is the power spectrum of the comoving 
curvature perturbation.  We comment how this 
might be calculated from the 
spectrum for the scalar field perturbations in the discussion section.

To follow even the modest approach and 
allow inhomogeneities in the scalar field, 
we must include a gradient term in
the matter Hamiltonian, ${\cal H}_{\phi}$.  
This term, strictly speaking, violates homogeneity but we will assume that the effect on the background spacetime is sufficiently small that it can be neglected.
Including this extra term, the matter part of
Eq. (\ref{HamSC}) becomes
\begin{eqnarray}
\label{hamphiscpert}
{\cal H}_{\phi+\delta \phi} &=&  \frac{1}{2} D(a)a^{-3} p_{\phi +\delta
\phi}^2 \nonumber \\
&~& + \frac{1}{2} \, a \, G(a)\, \delta^{ij} \partial_{i}(\phi + \delta \phi)
\partial_{j} (\phi + \delta \phi) \nonumber \\
&~& + a^3 V(\phi+\delta \phi) \,.
\end{eqnarray}
In this expression we have introduced a correction function $G(a)$, which 
we expect to arise due to the term $E^a_iE^b_i\partial_a \phi \partial_b \phi/ |\det E^c_j|^{-1/2}$ in Eq.~(\ref{HamPhi}) which involves inverse 
quantities, and must be regularized in a similar manner to the 
inverse volume term.  This term has so far not been calculated within LQC, and 
to attempt to do so here is beyond the scope of this note.  
Indeed in the previous study of perturbations in LQC this term was 
assumed to be unity.  The relevant regularization procedure for this 
term is closely connected to that for the inverse volume, and we anticipate 
it to have a similar form, in particular we assume that $G=1$ in the 
classical regime, but has a region for small values of the scale factor 
where $G \propto a^r$, where $r$ will depend on a new quantization parameter. 
In our study we will again assume, as in previous studies, that 
$r=0$. However, our method can easily be generalized to take account 
of a non zero $r$ and in the section VI we discuss how this would affect 
our results.

Using Eq.~(\ref{hamphiscpert}), the unperturbed 
Eqs.~(\ref{friedeq})--(\ref{raycheq}) are unaltered, but we have the additional
perturbation equation \be
\delta \phi'' = \left[-2 \frac{a'}{a}+\frac{D'}{D}\right] 
\delta\phi' + D \, \left[ \nabla^2- a^2\frac{d^2 V}{d\phi^2} \right]\delta\phi \,,
\label{pertfieldeq}
\ee
where a prime means differentiation with respect to conformal time $\tau$.
Using conformal time is helpful because it allows us to write 
Eq.~(\ref{pertfieldeq}) in the
particularly simple form
\be
u'' + \left( D k^2+m_{\rm eff}^2 \right) u = 0 \, ~,
\label{ueq}
\ee
where $u$ is defined as $u=aD^{-1/2}\delta\phi $,
and 
\be
\label{meff}
m_{\rm eff}^2  = - \frac{(a D^{-1/2})''}{a D^{-1/2}} +
a^2 D \frac{\partial^2 V}{\partial \phi^2}  \,,
\ee
is the effective mass of the field $u$.

In an equivalent approach, the scalar field equation in the
semiclassical LQC regime in the absence of metric perturbations,
can be derived from an effective action. In terms of conformal time, the action can be written as
\be
S = \int d \tau \, d^3 {\rm \bf x} {\cal L} = \int d\tau \, d^3{\rm \bf x} a^4
\left ( \frac{1}{2}\, \frac{\phi'^2}{D \,a^2} - V \right ) \,.
\ee
Adding the gradient term $-\delta^{ij} \partial_i \phi \partial_j \phi/2a$ to the quantity within brackets and
including a linear perturbation in the field around its background solution we find that the perturbed part of $S$ can be written as
\be
\label{uaction}
\delta S = \frac{1}{2}\int d\tau \, d^3 {\rm \bf x}
\left (u'^2 -D \,\delta^{ij}\partial_{i} u \, \partial_{j} u -
m_{\rm eff}^2 u^2 \right ) \,,
\ee
and when varied, this action also leads to Eq.~(\ref{ueq}).

\section{Power spectrum}
The action for $u$ (\ref{uaction}) 
is now formally equivalent to that of a scalar field with a
variable mass term, and a $D$ term multiplying the gradient part.
In order to calculate the spectrum of the perturbations,
produced during the super-inflation due to quantum fluctuations, we must
consider the field theory associated with the field $u$.

The momentum canonically conjugate to $u$ is given by
\be
\pi(\tau,x)=\frac{\partial {\cal L}}{\partial u'}=u'
\ee
The theory is then quantized by promoting $u$ and $\pi$ to
operators which satisfy the usual commutation relations.
We Fourier decompose $\hat u$ to give
\be
\label{uoperator}
\hat u= \int \frac{d^3k}{(2 \pi )^{3/2}} \left [ w_k(\tau ) \hat
a_{\rm \bf k} e^{i{\rm \bf k \cdot x}} + w_k^*(\tau ) \hat a_{\rm \bf k}^{\dagger} e^{-i{\rm \bf k \cdot x}} \right ]\, ~,
\ee
where $w_{k}$ are mode functions which satisfy the same equation as $u$
\be
\frac{d^2 w_k}{d \tau^2} + \left(D k^2 + m_{\rm eff}^2 \right) w_k = 0 \, ~.
\label{weq}
\ee
In order to have a well defined field theory, we must also 
ensure that $w_k$ is such that the creation and
annihilation operators, $\hat a_{\rm \bf k}^\dagger$ and $\hat a_{\rm \bf k}$, satisfy
the usual commutation relations for bosons. This means that $w_{k}$
must satisfy the Wronskian condition
\be
\label{wronskian}
w_k^* \frac{dw_k}{d \tau} - w_k \frac{dw_k^*}{d \tau} = -i \,.
\ee
In general, however, this condition does not give rise to a unique 
choice for $w_k$, instead  it allows a set of possible choices corresponding 
to a set of different Fock representations.
In the cosmological context a unique choice is 
normally determined by considering 
a limit in which the time dependence of the scale factor can 
be neglected, and hence where the physics ought to reduce to that of 
Minkowski space.  In this limit $w_k$ is normalized to select  
only the advanced solution. 
Once the initial condition is selected and the Wronskian condition met, a 
vacuum state is defined which is annihilated by all $\hat a_{\rm \bf k}$, 
such that $\hat a_{\rm \bf k} |0\rangle \,~= 0$.

The power spectrum of fluctuations about this vacuum state 
is defined by the vacuum expectation value such that
\be
\langle u_{\rm \bf k}u_{\rm \bf l}^* \rangle =\frac{2\pi^2}{k^3} {P_u} \delta^{(3)}({{\rm \bf k}-{\rm \bf l}})\,~,
\ee
where we have implicitly Fourier decomposed the field perturbation
$\delta \phi$, and
defined $u_{\bf k} = a \delta \phi_{\bf k}$.
Using Eq.~(\ref{uoperator}) we find
\be
\langle u_{\bf k} u_{\bf l}^*\rangle  =|w_k|^2\delta^{(3)}({{\bf k}-{\bf l}})\,~,
\ee
and hence that the power spectrum is given by
\be
\label{poweru}
{\cal P}_{u} = \frac{k^3}{2\pi^2}
|w_k|^2\, ~.
\ee

We now proceed to derive the form of the power spectra for the two 
cases under study.

\subsection{Massless field}

As we have seen, during the semiclassical phase
we have $D = D_* a^n$,
and we have power law growth with $a = A (-\tau)^p$ where $p = 2/(4-n)$.  
Inserting this into Eq.~(\ref{weq}) we obtain
\be
\label{w_kgeneral}
\frac{d^2 w_k }{ d \tau^2 } + \left (D_* \, A^n (-\tau)^{np} k^2 +
\frac{m_{\rm eff}^2 \tau^2}{\tau^2} \right )w_k = 0 \,.
\ee
For the massless case, using Eq.~(\ref{meff}), we find
\begin{equation}
m_{\rm eff}^2\, \tau^2 = -p(p-1) \,.
\end{equation}
The general solution admitted by Eq.~({\ref{w_kgeneral}) is,
\be
w_{k}(\tau) = c_{1} \sqrt{-\tau} \, J_{|\nu|} (x)
+ c_{2} \sqrt{-\tau} \, Y_{|\nu|} (x) \,,
\ee
where $J_{|\nu|}(x)$ and $Y_{|\nu|}(x)$ are Bessel functions of the first and second kind respectively and we have defined
\begin{equation}
\label{nudef}
\nu = - \frac{\sqrt{1-4 \, m_{\rm eff}^2 \, \tau^2}}{2+np}  \,,
\end{equation}
and
\begin{equation}
x = \alpha \, k \, (-\tau)^{(2+np)/2} = 
\left| \frac{2p}{2+np}\right| \, \frac{\sqrt{D}\,k}{aH} \,.
\end{equation}
with $\alpha = 2 \sqrt{D_* A^n}/|2+np|$ and $x > 0$.
We normalize this solution such that the Wronskian condition 
(\ref{wronskian}) is satisfied which in general gives
\begin{eqnarray}
\label{w_knormalised}
w_{k}(\tau) &=& \sqrt{\frac{\pi}{2|2+np|}} \,  \left( d_{1}\sqrt{-\tau} \,
H_{|\nu|}^{(1)}(x) \right.\nonumber \\ 
&~& + \left. d_{2} \sqrt{-\tau} \,H_{|\nu|}^{(2)}(x) \right) \,,
\end{eqnarray}
where $d_{1}$ and $d_{2}$ are constants subject to the condition $|d_{1}|^2-|d_{2}|^2=1$ and $H_{|\nu|}^{(1)}(x)$ and $H_{|\nu|}^{(2)}(x)$ are Hankel functions of the first and second kind, respectively. Moreover, the Hankel and Bessel functions are related through the expressions: 
$H_{|\nu|}^{(1)}(x) = J_{|\nu|}(x) + i \,Y_{|\nu|}(x)$ and  
$H_{|\nu|}^{(2)}(x) = J_{|\nu|}(x) - i \,Y_{|\nu|}(x)$.
We now consider the small wavelength limit in which the wavelength 
of the mode functions is far inside the cosmological horizon, and 
where we might expect a Minkowski form for the mode functions.
This limit corresponds to, $x \gg 1$, and the asymptomatic form
of Eq.~(\ref{w_knormalised}) is 
\begin{eqnarray}
w_k(\tau) &=& \frac{(-\tau)^{-np/4}}{\sqrt{|2+np|\, \alpha k}} \, 
\left(d_{1} \, \exp(i\alpha k (-\tau)^{(2+np)/2}) \right. \nonumber \\
&~& + \left. d_{2}  \, \exp(-i\alpha k(-\tau)^{(2+np)/2})  \right) \,.
\end{eqnarray}
In the standard inflationary scenario the analogous solution reduces to
two plane waves propagating in opposite directions in time, only the 
advanced solution is selected and $w_k(\tau) = e^{-ik\tau}/\sqrt{2k}$ 
in this limit.
In our case the solution only has the same form as flat spacetime 
when $n=0$, i.e. 
when the universe is classical. The two components to our solution 
however still represent advanced and retarded solutions, and by analogy we 
select only the advanced solution, this means we set $d_{1}=1$ and $d_{2}=0$. 
This can be justified by considering a mode whose wavelength remains 
well inside the cosmological horizon throughout the superinflationary evolution. 
Our normalization is then consistent with the Minkowski limit once super-inflation 
has ended. Moreover we note that ultimately our interest is in the 
$k$ dependence of the solution in the large wavelength limit, which
is unaltered by the normalization as long as the Wronskian condition 
is satisfied. 
That the solution does not reduce to the Minkowski limit however, 
already suggests that there are going to be 
clear differences in the evolution of perturbations with 
respect to the standard case  whenever a geometric correction 
to the kinetic term of the field occurs.

We can now look at the long wavelength limit of our properly normalized 
mode functions.  
The long wavelength limit is given by $k\ll 1$, and for a
specific finite time $\tau$ this corresponds to $x\ll 1$, and hence to
wavelengths well outside the effective horizon. In this limit we have
\begin{eqnarray}
\label{jsol}
J_{|\nu|}(x) &\rightarrow& \frac{1}{\Gamma(|\nu| +1)}
\left(\frac{x}{2}\right)^{|\nu|} \,, \\
\label{ysol}
Y_{|\nu|}(x) &\rightarrow& -\frac{\Gamma(|\nu|)}{\pi}
\left(\frac{x}{2}\right)^{-|\nu|} \,.
\end{eqnarray}
At this point a few comments are in order. As we have seen before, 
in the massless field case under study the growth power $p$ is negative, 
which means that the quantity $x \propto (-\tau)^{(2+np)/2} \propto (-\tau)^{2p}$ is an increasing 
function as $\tau \rightarrow 0$. Therefore, though the $Y_{|\nu|}(x)$ 
solution is the dominant one at early times, 
it is decreasing in nature and soon becomes sub dominant 
with respect to the increasing $J_{|\nu|}(x)$ solution. 
A related consequence of the 
growth power of $x$, $2p$, being negative is that as opposed to the standard 
inflationary scenario where the modes exit the effective 
horizon $1/aH$ during inflation, here, during super 
inflation driven by quantum effects, the modes {\it enter} 
the effective horizon given by $\sqrt{D}/aH$. 
Conversely, in the situation in which the universe is undergoing a collapsing evolution, modes eventually {\it exit} the horizon.
We will see that this not necessarily the case for the scaling solution.

While this is interesting, it raises a serious interpretational 
issue.  In standard inflation, the short wavelength limit is the 
same as the  $\tau\rightarrow -\infty$ limit, so all wavelengths can be considered 
to be small compared with the cosmological horizon at the earliest times. 
In this limit it is natural to assume that the small scale perturbations 
are governed by quantum mechanics and the normalization is performed in this 
limit.  As the expansion proceeds however the physical wavelength of the 
modes is increased, or equivalently the cosmological horizon size is 
decreased.  The modes are pushed outside the horizon 
as this behavior proceeds.  The modes effectively become 
classical, and the spectrum calculated in this limit can also be 
interpreted as a classical spectrum.  In the case at hand, however, this is 
no longer true, and it is not clear whether we can 
interpret the spectrum calculated on long wavelengths as a classical one. 

Taking this caveat on board, let us nevertheless 
proceed with the calculation. 
Using only the dominant part of Eq.~(\ref{w_knormalised}) 
with the help of Eq.~(\ref{ysol}), the power 
spectra Eq. (\ref{poweru}) reduces to
\begin{eqnarray}
\label{Pu}
{\cal P}_{u} &=& \frac{1}{4\pi} \left|\frac{p}{2+np}\right|^{1-2|\nu|} 
\left(\frac{\Gamma(|\nu|)}{\pi}\right)^2 \times \nonumber \\
&& \frac{a^2H^2}{D^{3/2}} \, \left(\frac{\sqrt{D}\,k}{aH}\right)^{3-2|\nu|}  
\propto  k^{3-2|\nu|} \, (-\tau)^{1-|\nu|(np+2)} \,, \nonumber \\
\end{eqnarray}
which for our massless field example (where $\nu = -n/8$ from Eq.~(\ref{nudef})), 
yields
\begin{equation}
{\cal P}_{u} \propto k^{3-2|\nu|} \, (-\tau)^{2(n-2)/(n-4)} \,.
\end{equation}
Then using ${\cal P}_{\phi} = D {\cal P}_{u}/a^2$, we find 
\begin{eqnarray}
\label{Pphi}
{\cal P}_{\phi} &\propto& \frac{H^2}{D^{1/2}} \left(\frac{\sqrt{D}\,k}{aH}\right)^{3-2|\nu|} \nonumber \\
&\propto&  k^{3-2|\nu|} \, (-\tau)^{1+p(n-2)-|\nu|(np+2)} \,,
\end{eqnarray}
which for the massless case, turns out to be time independent 
which tell us that the evolution of the scalar field perturbation 
is frozen on super horizon scales. We can also conclude that, for this case,
one obtains scale invariance of the scalar field perturbation 
only when $n = 12$.

\subsection{Scaling solution}

We can follow the same procedure for the self interaction potential 
with a scaling solution. In this case, using Eq.~(\ref{meff}) we obtain
\begin{equation}
m_{\rm eff}^2 \tau^2 = -2 + (3-2n)p +\frac{1}{2} (6+2n-n^2)p^2 \,,
\end{equation}
and for the quantity $\nu$ we have from Eq.~(\ref{nudef})
\begin{equation}
\nu = -\frac{\sqrt{9-12p+8np-12p^2-4p^2n+2n^2p^2}}{2+np} \,,
\end{equation}
where $p = -4/(n\beta+4)$. In this case we do not necessarily encounter the same behavior found in the massless case where the modes enter the horizon during the super inflationary phase. In fact, for $\beta > 2-4/n$ we have that $x$ is now 
decreasing as $\tau \rightarrow 0$. Hence, for these values of $\beta$, the dominant solution is always the $Y_{|\nu|}(x)$ function and the modes exit the effective horizon during the evolution.

In the limit of large $\beta$ we have that $p$ approaches zero and hence
we have $\nu \to -3/2$ which gives scale invariance. 
It is interesting to note that, in the limit of large $\beta$, 
the power spectrum is scale invariant regardless of $n$ (or the quantization parameter 
$l$). This is easy to understand because in this limit Eq.~(\ref{weq}) approaches
\be
w_k''+\tilde{k}^2w_k-\frac{2}{\tau^2} \,w_k = 0 \,,
\ee
with $\tilde{k}^2 = D_*A^n k^2$, 
which is of similar form 
to the analogous equation in the case of slow-roll inflation. 
Because the equation takes this form, the $x\ll1$ limit 
is identical to that for standard inflation and the 
Minkowski space limit is recovered, removing another 
conceptual problem.
The multiplicative factor in $\tilde{k}^2$ affects the normalization of the power spectrum but not the scale dependence, which is independent of $n$.  It is also 
interesting that no fine tuning of the $n$ (or $l$) parameter is required and 
that the solution is stable in the 
sense that $\beta \gg 1$ corresponds to background solutions which are stable 
to linear homogeneous perturbations \cite{Lidsey:2004uz}. This limit corresponds to the 
condition of a very steep potential which means that $a$, and hence $D$, are 
nearly constant despite $H$ varying. From Eq.~(\ref{meff}) we see that 
this means in this case the potential term is the most significant 
part of $m_{\rm eff}^2$.  

Using Eqs.~(\ref{Pu}) and (\ref{Pphi}) we see that in the limit of large $\beta$ we have for the power spectrum,
\begin{equation}
P_{\cal \phi} = \left(D_* A^{n+4}\right)^{-1/2} \, (2\pi\tau)^{-2} \,, 
\end{equation}
independent of $n$. 

We conclude that nearly scale invariance is a natural prediction 
of the LQC universe sourced by a 
steep potential of the form given by Eq. (\ref{potential}).  
As will be discussed below, we must not over emphasize this result
as it may change if metric perturbations are significant.

There is, however, an additional 
solution for a particular 
fine tuned value of $\beta$ which gives $\nu=3/2$ and hence also scale 
invariance.  Indeed, as we decrease the value of $\beta$, we see that at 
$np = -2$ or $\beta = (2n-4)/n$, the value of $\nu$ blows up to $-\infty$ and switches sign. As $\beta$ approaches zero, $p$ approaches $-1$ and consequently $ \nu \rightarrow \sqrt{9-12n+2n^2}/(n-2)$ which is always between 
$\sqrt{2} < \nu < 3/2$, therefore, $\nu$ must cross the value $\nu = 3/2$ at small $\beta$. A scale invariant power spectrum is therefore possible for small $\beta$ but subject to a severe fine tuning.

\section{Discussion}
In this article we have computed the power spectrum of the scalar field perturbations for two distinct situations. First we have considered the dynamics of a massless scalar field. We found the interesting behavior that the modes of the scalar field perturbations enter the effective horizon during the superinflationary phase in clear contrast with the evolution in standard slow-roll inflation. Scale invariance is possible in this case but at the cost of the fine tuning of the quantization parameter $l$. However we note 
that the required value $n = 12$ ($l=2/3$) is not one of the values which 
are favored by consideration of the full theory.
 
An interesting question is whether allowing the parameter function $G$ to vary from unity would modify the phenomenology.  It is easy to see that the effect 
of the function $G$ on the $k$ dependence, is to change Eq.~(\ref{nudef}) 
such that $\nu=-(1-4 m_{\rm eff}^2\tau^2)^{1/2}/(2+np+rp)$, while $m_{\rm eff}$ 
remains unaltered. Hence there is an extra degree of freedom in this case, and 
it would be interesting to investigate whether scale invariance can be 
achieved  without moving away from the preferred values of 
quantization parameters using this freedom.  

Before leaving the massless 
case it is also worth noting that even in the limit $l=0$, which represents
exponential expansion $\dot{H}=0$, we do not find a scale variant spectrum and 
we still have the problem of modes entering the horizon. This is in stark contrast with earlier work \cite{Hossain:2004wm} which, at least in part, motivated 
our investigation.  Indeed within our calculation even if we had imposed by hand 
that the background evolution was exponential (i.e. fixing $p = -1$ but leaving $n$ free), 
we would have obtained scale invariance provided that $n = 0$ or $n = 12/5$. Again, in contrast with the previous study.

The fine tuning displayed above 
is evaded in the second situation we investigated, that of the scaling solution. By including a self interaction potential, we gain a degree of freedom that can be used to set the region of parameter space that offers the desired features of the inflationary scenario i.e. modes exiting the horizon and near scale invariant power spectrum of perturbations.  Moreover it is clear that allowing the $G$ function to vary will not affect this limiting behavior 
since the $p\rightarrow 0$ limit ensures $G$ will be close to a constant.

At this point in the discussion it is useful to say something about when we 
might expect our calculation to be accurate, and how it could be 
applied to derive the comoving curvature perturbation. This quantity is 
useful since under very general circumstances it is conserved on super-horizon 
scales.  Hence its spectrum calculated as modes leave the horizon during 
inflation is equal to the spectrum as these modes re-enter at a later time, 
when they account for the formation of cosmic structure.
In our calculation we have simply calculated the scalar field spectrum during 
super-inflation, however, we would like to convert this result into the comoving 
curvature spectrum.  To make this conversion we must pick a gauge in 
which we assume that the metric perturbations in the scalar field 
equation are sub dominant to the scalar field perturbations, as in this 
gauge our calculation will be accurate. If this 
is the spatially flat gauge then the conversion to the curvature 
perturbation would simply be given by $P_{\cal R}=({\cal H}/\phi')^2{P_\phi}$. To convert the spectrum in this way is the procedure used in 
Ref. \cite{Liddle:2000cg} for standard inflation. 
However, until the full equations of gravitational and matter perturbations 
are known we will not know in what gauges (if any) the metric perturbations 
can be ignored, and hence how to convert our spectrum to the 
spectrum of curvature perturbations.   

In addition to the inclusion of the background perturbations into our calculation, 
a further way to improve its accuracy consists of relaxing the 
assumption that $a\ll a_*$, and hence by using the full form of the function $D(a)$, 
rather than its asymptotic approximation. This would require the mode 
functions to be solved numerically, and it would be interesting 
to compare this approach with the analytic results derived here. 

Nonetheless, our aim  was not to establish a robust prediction for the 
spectral index from LQC inflation, 
but rather to illustrate three important points. First, subject
to the approximations we have to make, a scale invariant spectrum
is possible for LQC inflation even when the equation of state differs
from $w\approx-1$. This is a great surprise considering
our experience from standard inflation.  Second, even
when the universe is super-inflating the Fourier space
modes of the scalar field perturbation
are not necessarily pushed outside a suitably defined horizon. 
We encountered this type of behavior in the massless case. This again
is unexpected considering standard inflation.
Third, just like in
slow-roll inflation, when calculated with the standard
techniques, there
is considerable freedom 
in the value of the spectral index from LQC inflation.
This freedom is related to the form of the
potential and, uniquely to LQC, it is also dependent on the choice
of quantization ambiguities.  In particular we expect that, if we consider 
potentials other than those that generate a scaling solution, 
the spectral index may be greater or less than unity.

We argue, therefore, that our calculation is an important towards understanding the phenomenology of LQC. In particular
it highlights these three features which
are likely to carry over to a full analysis
including metric perturbations, and which deserve considerable
attention.
The calculation is also important given the likely 
complexity of the full perturbed equations.  Once the full equations 
are known we will be able to determine when our 
calculation ought to provide an accurate answer, and in these 
cases it will provide a useful check on any 
spectrum of perturbations calculated using the full equations. 
As we have seen there are many subtleties in any calculation of a 
perturbation spectrum. It may even 
be that the approach of using the full equations to determine 
when background perturbations 
can be ignored and performing the calculation given here, is the only 
case in which the spectrum of perturbations from inflation in LQC can be determined 
using the standard techniques. 

\begin{acknowledgments}
The authors thank Martin Bojowald, Golam Hossain, James Lidsey and Reza Tavakol for 
discussions and comments on the manuscript. The work of DJM was supported by PPARC and the work of NJN was supported by DOE grant DE--FG02--94ER--40823.
\end{acknowledgments}

\end{document}